\documentstyle[aps,prl,twocolumn,epsfig,floats]{revtex}

	\newcommand{\Ref}[1]{(\ref{#1})} 
	\newcommand{\dst}{\mbox{$\displaystyle$}}

\begin{document}


	\title{Classical Advection of Guiding Centers in a Random Magnetic 
	Field}

	\author{L.~Zielinski$^1$, K.~Chaltikian$^1$, K.~Birnbaum$^1$, 
	C.~M.~Marcus$^1$, K.~Campman$^2$ and A.~C.~Gossard$^2$} 
	\address{$^1$Department of Physics, Stanford University, Stanford, 
	CA 94305-4060} \address{$^2$Materials Deparment, University of 
	California at Santa Barbara, Santa Barbara, CA 93106}
	\date{\today}
	\maketitle
\begin{abstract} We investigate theoretically and experimentally 
classical advective transport in a 2D electron gas in a random 
magnetic field.  For uniform external perpendicular magnetic fields 
large compared to the random field we observe a strong {\em 
enhancement} of conductance compared to the ordinary Drude value.  
This can be understood as resulting from advection of cyclotron 
guiding centers.  For low disorder this enhancement shows non-trivial 
scaling as a function of scattering time, with consistency between 
theory and experiment.  \end{abstract}
	\pacs{PACS numbers: 72.10.-d, 72.10.Bg, 71.55.Jv} 


Transport in two-dimensional (2D) electron systems in a spatially 
random magnetic field (RMF) has generated great theoretical 
\cite{theory,HedSmith} and experimental interest 
\cite{experiment,mancoff,mancoff2} in recent years, and is now 
understood to be distinctly different from transport in systems with 
ordinary potential disorder.  Experimentally, these systems have been 
realized using high-mobility heterostructure materials and overlayers 
of superconductors or ferromagnets \cite{experiment,mancoff,mancoff2} 
and may be related to quantum Hall transport around filling factor 
$1/2$ \cite{theory}.

In the presence of a strong external uniform magnetic field and 
negligible potential disorder, electrons in a RMF move in spiraling 
cyclotron orbits with guiding centers moving along the contours of 
constant magnetic field \cite{Northrop}.  Because contours are 
generically closed \cite{isichenko1} electrons confined to contours do 
not contribute to total conductance in the limit of zero cyclotron 
radius.  A similar situation arises in the case of electron gas moving 
in a long-range potential and strong uniform magnetic field, where the 
guiding centers move along the contours of constant potential, as 
recently discussed by Fogler {\em et al.} \cite{FogShk}.  In that 
case, authors of \cite{FogShk} showed that a finite cyclotron radius 
allowing the electrons to ``jump'' between different contours only 
leads to an exponentially small conductance.  In general, scattering 
from short-range potential disorder increases the probability of such 
jumps, significantly enhancing the conductance via the same mechanism: 
by freeing those electrons which would otherwise be trapped on closed 
contours of the RMF or long-range potential disorder.  That type of 
transport may be called advective, in analogy with the well-studied 
problem in fluid dynamics where the transport of a tracer particle due 
to molecular diffusion becomes greatly enhanced due to the flow of the 
fluid \cite{fluid-dyn}.

A useful measure to analyze transport in a system with both RMF and 
ordinary potential disorder is the ratio $R = 
\sigma_{xx}^{\rm\scriptscriptstyle RMF}/\sigma_{xx}$ of the 
longitudinal conductances with and without the RMF. In this Letter we 
present a novel analysis of advective transport in a RMF, leading to 
non-trivial scaling of the ratio $R$ with the scattering time of the 
ordinary potential.  We then compare theoretical predictions to 
experiments in a high-mobility 2D electron gas (2DEG) with a spatially 
random high-field magnet on the surface and controllable potential 
disorder.


We begin by discussing the classical statistical theory of electron 
motion in an inhomogeneous perpendicular magnetic field, $B({\bf r})$.  
The following notation is used throughout the paper: $\omega_0$ and 
$\delta\omega$ represent respectively the average and standard 
deviation of the random cyclotron frequency $\omega_c({\bf r}) = {e 
B({\bf r})\over m^*c}$, $\xi_B$ and $\xi_P$ are the experimentally 
determined correlation lengths of the RMF and ordinary potential 
$V({\bf r})$ respectively.  The strengths and spatial scales of the 
two independent disorders relevant to the experiment satisfy the 
inequalities $\xi_P \lesssim \dst v_{\rm\scriptscriptstyle 
F}/\max{(\omega_0,\delta\omega)} \ll \xi_B$, $V_{\rm\scriptstyle rms} 
\ll E_F$, where $v_{\rm \scriptscriptstyle F}$ is the Fermi velocity 
and $E_F$ is the Fermi energy, and it is these limits that we consider 
in this Letter.  Note that the first inequality implies only that the 
random magnetic field varies slowly with position; no assumption about 
its strength is made.

To study transport in this mixed-disorder regime, we take advantage of 
the assumed separation of disorder length scales.  Following the 
analysis of Ref.  \cite{HedSmith}, magnetic randomness is treated as a 
Lorentz force term in the left hand side of the Boltzmann equation, 
while potential randomness leads to diffusion, characterized by a 
transport time $\tau_{\rm\scriptstyle tr}$.  Charge conservation then 
implies a diffusion equation for the fluctuating part of the particle 
density, 
\begin{equation} {\partial n\over\partial t} = 
{D_0\over{1\!+\!\beta^2}}\mbox{\boldmath $\nabla$}^2 n + 
{D_0\over{1\!+\!\beta^2}}\left( 
{\beta^2\!-\!1\over\beta^2\!+\!1}\mbox{\boldmath 
$\nabla$}_{\perp}\beta - {2\beta\mbox{\boldmath 
$\nabla$}\beta\over{\beta^2\!+\!1}}\right)\!\cdot\!  \mbox{\boldmath 
$\nabla$} n, 
\label{diffus}
\end{equation} 
where $\beta({\bf r})=\omega_c({\bf 
r})\tau_{\rm\scriptstyle tr}$ is the local dimensionless cyclotron 
parameter, $D_0 = \frac{1}{2}v_{\rm\scriptscriptstyle 
F}^2\tau_{\rm\scriptstyle tr}$ is the ordinary diffusion coefficient 
and $\mbox{\boldmath$\nabla$}_{\perp} = \left(\partial/\partial y, 
-\partial/\partial x\right)$ \cite{karen-thesis}.  We will refer to 
the first and second terms in the r.h.s. as {\em diffusive} and {\em 
advective} terms respectively.

Equation \Ref{diffus} generalizes the standard advection-diffusion 
problem studied in fluid dynamics and plasma physics \cite{isichenko1} 
described by the equation
\begin{equation}\label{ad-dif} {\partial n\over\partial t} = D_* 
\mbox{\boldmath $\nabla$}^2n + {\bf u}\cdot\mbox{\boldmath $\nabla$}n.
\end{equation}
In the long-time, long-wavelength limit, solutions to Eq.  
\Ref{ad-dif} are known to converge to solutions of the usual diffusion 
equation \cite{PapVar,McL}, $$ {\partial n\over\partial t} = 
D_{\rm\scriptstyle eff}\mbox{\boldmath $\nabla$}^2 n, $$ with an 
effective diffusion constant, $D_{\rm\scriptstyle eff}$.  In the 
standard advection-diffusion problem, Eq.\Ref{ad-dif}, one 
distinguishes two regimes: the {\em diffusive} regime, in which the 
first (diffusive) term on the right hand side of Eq.  \Ref{ad-dif} 
dominates and advection may be ingnored, and the {\em advective} 
regime, in which the advective term makes a significant contribution 
to the transport coefficients.  The dimensionless parameter 
characterizing the relative strength of advection over diffusion is 
the so-called {\em Pecl\'{e}t number} $P = \xi u_0/D_*$, where $\xi$ 
and $u_0$ are respectively the characteristic spatial scale and 
amplitude of the random velocity field ${\bf u(r)}$.

Now, returning to our problem, Eq.~\Ref{diffus}, the Pecl\'{e}t number 
in this case is approximately given by the RMF cyclotron parameter, 
$P\approx \delta\omega\, \tau_{\rm\scriptstyle tr}$ and the two 
corresponding transport regimes are defined as follows:

\noindent {\bf (i)} Strong disorder, when the inequality 
$\tau_{\rm\scriptstyle tr}^{-1} \gg \max(\delta\omega,\omega_0)$ is 
obeyed.  This corresponds to the diffusive regime.  Here the advective 
term is negligible and the effective diffusion constant is 
approximated by $D_{\rm\scriptstyle eff}/D_0\approx \dst 
\left\langle{1\over 1+\beta^2}\right\rangle.$

\vskip 0.2cm \noindent {\bf (ii)} Weak disorder and strong magnetic 
field, when the inequality $\tau_{\rm\scriptstyle tr}^{-1} \ll 
\delta\omega \lesssim \omega_0$ is obeyed.  This corresponds to the 
advective regime.  In this regime Eq.\Ref{diffus} is well approximated 
by the form of Eq.~\Ref{ad-dif} with $D_* = D_0\left\langle{1\over 
1+\beta^2}\right\rangle$ and ${\bf u} = D_*{\bf\nabla}_{\perp}\beta$.  
The dependence of $D_{\rm\scriptstyle eff}$ on the bare diffusion 
constant $D_*$ in this regime, characterized by large Pecl\'{e}t 
numbers, ($P\approx \delta\omega\tau_{\rm\scriptstyle tr}\gg 1$) has 
been a subject of several theoretical and numerical studies over the 
past decade \cite{DrDyk,isichenko2,AvMd,papanic-rand,isichenko1}.  
Heuristic arguments \cite{isichenko2} and non-local variational 
principles \cite{papanic-rand} lead to the expression 
$D_{\rm\scriptstyle eff}/D_*\approx a(\tau_{\rm\scriptstyle tr}) 
P^{\alpha_0}$, where $a(\tau_{\rm\scriptstyle tr})= 
{1+(\omega_0\tau_{\rm\scriptstyle tr})^2\over 
(\delta\omega\tau_{\rm\scriptstyle tr})^{2\alpha_0} + 
(\omega_0\tau_{\rm\scriptstyle tr})^2}$ \cite{karen-thesis}.  In the 
limit $\delta\omega\,\tau_{\rm\scriptstyle tr}\gg 1$, the prefactor 
$a(\tau_{\rm\scriptstyle tr})\to 1$, giving the simple scaling form 
\begin{equation} \label{10/13} 
D_{\rm\scriptstyle eff}/D_* \approx 
\dst P^{\alpha_0}, \end{equation} where \begin{equation} \label{power} 
\alpha_0 = \frac{1+\nu d_f}{2+\nu d_f}  
\end{equation} 
is a constant that depends on the geometry and statistical properties 
of the random surface.  In Eq.\Ref{power}, $\nu$ is the critical 
exponent for the correlation length of the level contours of $B({\bf 
r})$, and $d_f$ is the fractal dimension of these contours 
\cite{isichenko1}.  If $B({\bf r})$ has Gaussian statistics and is 
short-range correlated, then $\nu = {4\over 3}$ and $d_f={7\over 4}$ 
\cite{SalDup}, giving $\alpha_0 = {10\over 13}\approx 0.77$ 
\cite{isichenko2}.  For non-Gaussian statistics of $B({\bf r})$, the 
value of $\nu$ depends on the distribution of saddle-point 
heights\cite{gamma}.  If $B({\bf r})$ is doubly periodic, for 
instance, then $\nu = d_f = 0$, giving $\alpha_0 = {1\over 2}$, a 
result known from boundary layer considerations \cite{Childress}.  The 
scaling relation, Eq.\Ref{10/13}, is the main signature of advective 
transport that we investigate experimentally, as described below.
\begin{figure}[bth] 
	\epsfxsize=\columnwidth 
    \epsfbox{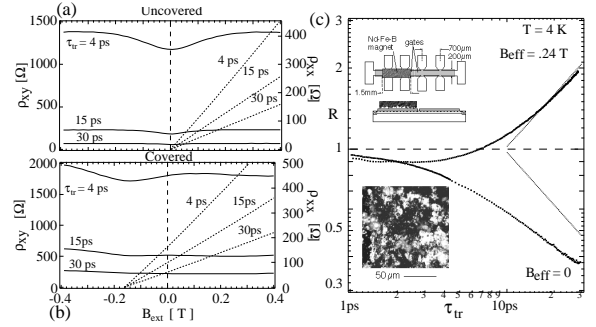} 
    \vspace{5 mm} 
	\caption{ (a) Hall (left axis) and longitudinal (right axis) 
	resistivities for the uncovered half of the Hall bar at three 
	different values of $\tau_{\rm\scriptstyle tr}$ vs.  external 
	magnetic field, $B_{\rm\scriptstyle ext}$.  Note that for the 
	uncovered half $B_{\rm\scriptstyle eff} = B_{\rm\scriptstyle 
	ext}$.  (b) Corresponding graphs for the covered half.  Note the 
	shift of the effective magnetic field, $B_{\rm\scriptstyle eff} = 
	B_{\rm\scriptstyle ext} - 0.15T$.  Zero of $B_{\rm\scriptstyle 
	eff}$ is by definition the point of intersection of Hall slopes.  
	(c) The ratio $R=\sigma^{\rm\scriptscriptstyle 
	RMF}_{xx}/\sigma_{xx}$ plotted for two representative values of 
	$B_{\rm\scriptstyle eff}$ showing the suppression of the RMF 
	conductance at $B_{\rm\scriptstyle eff}=0$ and the strong 
	enhancement due to advection at high field.  Dotted lines show the 
	theoretical expression $R= a(\tau_{\rm\scriptstyle tr}) 
	P^{\alpha_0}$.  In the limit $\tau_{\rm\scriptstyle tr}\to\infty$, 
	$R= P^{\alpha_0}$ in the advective regime ($B_{\rm\scriptstyle 
	eff} = 0.24T$) and $R= P^{-\alpha_0}$ in the diffusive regime 
	($B_{\rm\scriptstyle eff} = 0$).  In both cases the values 
	$\alpha_0=0.65$ and $B_{\rm\scriptstyle rms}=0.04T$ giving the 
	best fit were used.  Upper inset: Schematic of the device, 
	including gates and attached magnet.  Lower inset: Optical 
	micrograph of the Nd-Fe-B surface indicating roughness on the 
    $\sim 20\mu m$ scale.} \label{fig1} 
\end{figure}
The experimental system consists of a $200\mu m$ wide by $3mm$ long 
Hall bar fabricated from a high-mobility GaAs/AlGaAs heterostructure 
(see Fig.~\ref{fig1}c, inset).  Sheet density and scattering times are 
controlled by applying voltages in the range $V_{\rm\scriptstyle gate} 
= -1.2$ to $+0.7 V$ via two independent gates thermally evaporated 
onto one half of each structure (gate material is $100\AA$ Cr followed 
by $2500\AA$ Au).  To generate the random magnetic field, a 
neodymium-iron-boron (Nd-Fe-B) permanent magnet ($300\mu m$ wide, 
$200\mu m$ tall, $1.5mm$ long) was affixed with polymethyl 
methacrylate (PMMA) to the Hall bar above the gate with the easy 
magnetization axis of the sintered material perpendicular to the 
surface \cite {mancoff,mancoff2}.  The Nd-Fe-B magnet is attached in a 
demagnetized state, but after cooling the sample and ramping an 
external perpendicular magnetic field $B_{\rm\scriptstyle ext}$ to 
$6T$, the material becomes permanently magnetized, creating a field of 
$B_0\sim 0.15 T$ at the electron gas as measured from the Hall 
resistance offset, seen as the intersection point of the Hall 
resistances in Fig.  \ref{fig1}b.  The effective average field 
$B_{\rm\scriptstyle eff}$ felt by the electrons is the difference 
between the applied field and this constant offset, 
$B_{\rm\scriptstyle eff} = B_{\rm\scriptstyle ext}-B_0$.

The rough surface of the Nd-Fe-B (Fig.~\ref{fig1}c, lower inset) 
generates a spatially random magnetic field at the 2DEG with standard 
deviation $\delta B\sim 0.04T$ (measured by a best fit low-field 
magnetoresistance \cite{mancoff}), and characteristic length scale 
$\xi_B\approx 20\mu m$.  The permanent magnetization is insensitive to 
changes in $B_{\rm\scriptstyle ext}$ in the range $\sim -0.5T$ to 
$0.5T$ of interest in the experiment, and little hysteresis is 
observed in this range following initial magnetization.  Longitudinal 
and Hall resistances were measured at 4.2K on both the end of the Hall 
bar under the magnet (``covered'') and the end without the magnet 
(``uncovered'') using standard ac lock-in techniques at $3Hz$ with a 
current bias of $100nA$.  Following the transport measurements in the 
RMF, the Nd-Fe-B magnet was taken off and the measurements repeated on 
the previously covered side.  We note that at these temperatures and 
magnetic fields quantum effects in the form of Shubnikov-de-Haas 
oscillations are not observed \cite{mancoff2}.  Conductivities with 
and without the RMF, $\sigma^{\rm\scriptscriptstyle RMF}_{xx}$ and 
$\sigma_{xx}$, were then computed from the measured $\rho_{xx}$'s and 
$\rho_{xy}$'s as a function of gate voltage.  The non-RMF data are 
based on the runs after removing the magnet, with the uncovered half 
of the sample serving as a control.  The dependence of 
$\tau_{\rm\scriptstyle tr}$ on $V_{\rm\scriptstyle gate}$ was found to 
be repeatable for a particular sample through multiple thermal 
cyclings, allowing $V_{\rm\scriptstyle gate}$ to serve as a reliable 
and repeatable `knob' controlling $\tau_{\rm\scriptstyle tr}$.

\begin{figure}[bth] 
   \epsfxsize=\columnwidth 
   \epsfbox{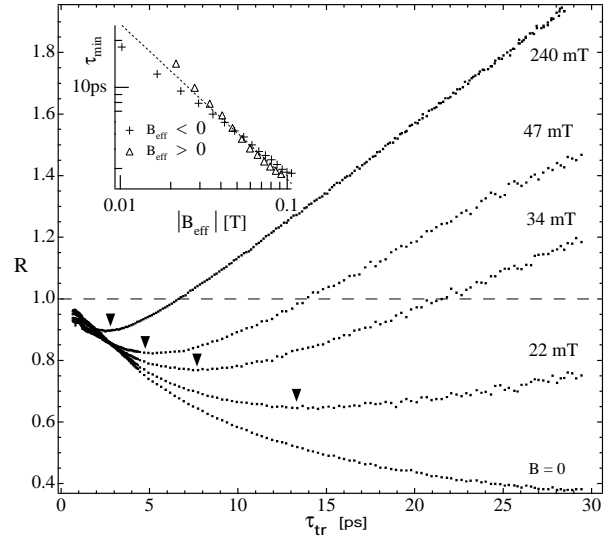} 
   \vspace{5 mm} 
   \caption{The ratio 
		$R=\sigma^{\rm\scriptscriptstyle RMF}_{xx}/\sigma_{xx}$ 
		measured at different values of $B_{\rm\scriptstyle eff}$ 
		showing the gradual shift of the transition region between the 
		diffusive and the advective regimes with increasing 
		$B_{\rm\scriptstyle eff}$.  The location of the minima of $R$, 
		$\tau_{\rm\scriptstyle min}$, indicated by triangles, is 
		plotted in the inset as a function of $B_{\rm\scriptstyle 
		eff}$ showing a power law dependence $\tau_{\rm\scriptstyle 
		min}\propto B_{\rm\scriptstyle eff}^{-1}$.}
		\label{fig2} 
\end{figure}

\begin{figure}[bth] 
  \epsfxsize=\columnwidth \epsfbox{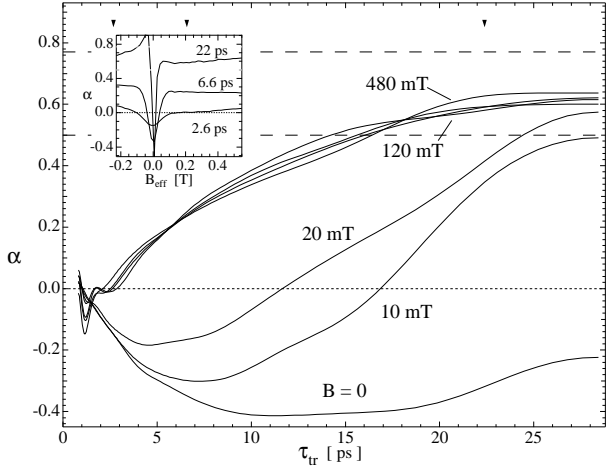} \vspace{5 mm} 
  \caption{ Scaling exponent $\alpha = 
  d\ln(R)/d\ln(\tau_{\rm\scriptstyle tr})$ as a function of 
  $\tau_{\rm\scriptstyle tr}$ at different values of 
  $B_{\rm\scriptstyle eff}$.  The saturation at $\alpha \approx 0.65$ 
  for $B_{\rm\scriptstyle eff} > 120 mT$ and $\tau_{\rm\scriptstyle 
  tr} > 20 ps$ is a signature of the advective regime and corresponds 
  to a power law $R\propto \tau_{\rm\scriptstyle tr}^{\alpha}$.  The 
  two horizontal dashed lines correspond to the two limiting cases of 
  periodic ($\alpha = 0.5$) and gaussian disordered $B({\bf r})$ 
  ($\alpha\approx 0.77$).  One expects $0.5 < \alpha < 0.77$ for any 
  physical system.  Inset shows $\alpha$ as a function of 
  $B_{\rm\scriptstyle eff}$ at three values of $\tau_{\rm\scriptstyle 
  tr}$ (marked by filled triangles in the main figure.  The topmost 
  curve at $\tau_{\rm\scriptstyle tr} = 22ps$ asymptotically 
  approaches $0.65$ for high values of $B_{\rm\scriptstyle eff}$.  For 
  larger values of $\tau_{\rm\scriptstyle tr}$, $\alpha$ remains 
  constant at $0.65$ for $B_{\rm\scriptstyle eff} >0.1T$.}
	\label{fig3} 
\end{figure}

Two samples with different ranges of transport elastic scattering time 
$\tau_{\rm\scriptstyle tr}(V_{\rm\scriptstyle gate})$ are reported.  
In sample 1, $\tau_{\rm\scriptstyle tr}$ ranged from $0.1$ to $10ps$, 
corresponding to ranges $n=7.9\!\times\!  10^{15} - 1.2\!\times\!  
10^{16}m^{-2}$ and $\mu= 0.23 - 26 m^2/(V\, s)$.  In sample 2, 
$\tau_{\rm\scriptstyle tr}$ ranged from $0.5$ to $30ps$, corresponding 
to ranges $n=9.9\!\times\!  10^{14} - 3.8\!\times\!10^{15}cm^{-2}$ and 
$\mu= 2.2 - 78 m^2/(V\, s)$.  We will concentrate on data from the 
``cleaner'' sample 2 since the advective regime is reached more easily 
for larger $\tau_{\rm\scriptstyle tr}$.
Figure \ref{fig1}c shows the ratio $R=\sigma_{xx}^{\rm\scriptstyle 
RMF}/\sigma_{xx}$ as a function of $\tau_{\rm\scriptstyle tr}$ at 
effective magnetic field $B_{\rm\scriptstyle eff} = 0$ and 
$B_{\rm\scriptstyle eff} = 0.24T$.  The ratio $R$ is equivalent to the 
ratio $D_{\rm\scriptstyle eff}/D_*$ and so is expected to scale with 
$\tau_{\rm\scriptstyle tr}$ as in Eq.~\Ref{10/13} in the advective 
regime.  Figures \ref{fig1}c illustrates the key difference between 
the advective and diffusive regimes: At low $B_{\rm\scriptstyle eff}$ 
-- the diffusive regime -- the RMF acts to increase the total disorder, 
so that as $\tau_{\rm\scriptstyle tr}$ is increased (potential 
scattering reduced),the RMF becomes the dominant source of scattering, 
leading to a {\em decreasing} R with increasing $\tau_{\rm\scriptstyle 
tr}$.  On the other hand, for larger $B_{\rm\scriptstyle eff}$ the 
advective regime is reached and the RMF causes advection of guiding 
centers, leading to an {\em increasing} R with increasing 
$\tau_{\rm\scriptstyle tr}$, as predicted by Eq.  \Ref{10/13}.  Figure 
\ref{fig2} shows in greater detail how this transition from the 
diffusive to the advective regime depends on $B_{\rm\scriptstyle 
eff}$.  The transition is indicated by a crossover from decreasing to 
increasing $R$ with increasing $\tau_{\rm\scriptstyle tr}$, marked by 
a triangle below each trace in Fig.  \ref{fig2}.  For higher 
$B_{\rm\scriptstyle eff}$ the crossover region moves to lower values 
of $\tau_{\rm\scriptstyle tr}$ with the minimum of 
$R(\tau_{\rm\scriptstyle tr})$ depending on $B_{\rm\scriptstyle eff}$, 
according to $\omega_0\tau_{\rm\scriptstyle min}\sim 6$ (see 
Fig.~\ref{fig2}, inset).
To look for scaling in the advective regime, $\tau_{\rm\scriptstyle 
tr}>\tau_{\rm\scriptstyle min}$, we define a scaling exponent 
\begin{equation} \label{alpha-def} \alpha(B_{\rm\scriptstyle 
eff},\tau_{\rm\scriptstyle tr})\ =\ {d\ln{(R)}\over 
d\ln{(\tau_{\rm\scriptstyle tr})}} \end{equation} comparable to 
$\alpha_0$ in Eq.~\Ref{10/13}.  Figure~\ref{fig3} shows the measured 
$\alpha$ as a function of $\tau_{\rm\scriptstyle tr}$ at different 
values of $B_{\rm\scriptstyle eff}$ and in the inset as a function of 
$B_{\rm\scriptstyle eff}$ at fixed $\tau_{\rm\scriptstyle tr}$.  We 
see that only for $B_{\rm\scriptstyle eff} \sim 0$ (in the 
high-mobility sample for $B_{\rm\scriptstyle eff}\lesssim 0.01T$) does 
$\alpha$ remain negative for all values of $\tau_{\rm\scriptstyle 
tr}$.  This behavior indicates that $R$ decreases monotonically around 
$B_{\rm\scriptstyle eff} \sim 0$ as already seen in Fig.~\ref{fig1}c).  
As we increase $B_{\rm\scriptstyle eff}$, $\alpha$ levels off for 
large values of $\tau_{\rm\scriptstyle tr}$, which we interpret as 
having entered the advective regime of \Ref{10/13}.  A characteristic 
feature of $\alpha$ as a function of $B_{\rm\scriptstyle eff}$ (inset, 
Fig.~\ref{fig3}) is the dip near zero that becomes deeper for larger 
$\tau_{\rm\scriptstyle tr}$.  This is expected from our prediction 
that $\alpha(0,\tau_{\rm\scriptstyle tr})$ should be monotonically 
decreasing with increasing $\tau_{\rm\scriptstyle tr}$.

Figure~\ref{fig3} indicates that the advective regime is first reached 
at $\tau_{\rm\scriptstyle tr} = 22ps$ ($P = 5.72$), at high magnetic 
fields, where $\alpha(B_{\rm\scriptstyle eff},22ps)\to 0.65$.  The 
$\tau_{\rm\scriptstyle tr} = 22ps$ data, shown as the topmost curve in 
the inset of Fig.~\ref{fig3}, emphasizes the asymptotic nature of the 
evolution of $\alpha$.  At higher $\tau_{\rm\scriptstyle tr}$, for 
$B_{\rm\scriptstyle eff} \gtrsim 0.1T$, $\alpha(B_{\rm\scriptstyle 
eff},\tau_{\rm\scriptstyle tr})$ is virtually constant at $0.65$.  
This {\em constant} value of $\alpha$, independent of 
$\tau_{\rm\scriptstyle tr}$ and $B_{\rm\scriptstyle eff}$ is the 
experimental signature of the expected power law, $R \propto 
\tau_{\rm\scriptstyle tr}^{\alpha}$.  The measured scaling exponent 
$\alpha = 0.65$ is within the range $1/2 < \alpha < 10/13$ expected 
from the present theory.  The specific attributes of the experiment 
which lead to the particular value 0.65 are not understood at this 
point, however it is known that deviations from gaussian fluctuations 
of $B({\bf r})$ will change the scaling of advective transport within 
the range allowed by Eq.~\Ref{power}.


In summary, we have investigated classical advection of guiding center 
motion of disordered 2DEG in a RMF, and found theoretically a novel 
scaling of the ratio $R$ of diffusion constants or conductivities with 
and without the RMF as a function of $\tau_{\rm\scriptstyle tr}$.  
Experiments confirm the expected power law scaling, and give a scaling 
exponent consistent with theory.

We thank S.~C.~Zhang and G.~C.~Papanicolaou for useful discussion.  We 
acknowledge support at Stanford from the ONR under Grant 
N00014-94-1-0622, the Army Research Office under grant 
DAAH04-95-1-0331, the NSF-NYI program (C.M.M.), the NSF under grant 
DMR-9522915 (K.C.), the NSF Center for Materials Research, and the URO 
program (L.Z.) and Terman Fellowship at Stanford.  We also acknowledge 
support at UCSB by the AFOSR under Grant F49620-94-1-0158 and QUEST.


\begin{references}
\bibitem{theory} 
	 V.~Kalmeyer and S.~C.~Zhang, {\em Phys.  Rev.} {\bf B46}, 9889 
	 (1992); B.~I.~Halperin, P.~A.~Lee and N.~Read, {\em Phys.  Rev.} 
	 {\bf B47}, 7312 (1993); V.~Kalmeyer, D.~Wei, D.~Arovas and 
	 S.~C.~Zhang, {\em Phys.  Rev.} {\bf B48}, 11095 
	 (1993);B.~L.~Altshuler and L.~B.~Ioffe, {\em Phys.  Rev.  Lett.} 
	 {\bf 69}, 2979 (1992); D.~V.~Khveshchenko and S.~V.~Meshkov, {\em 
	 Phys.  Rev.} {\bf B47}, 12051 (1993); A.~G.~Aronov, A.~D.~Mirlin 
	 and P.~Wolfle, {\em Phys.  Rev.} {\bf B49}, 16609 (1994); 
	 K.~Chaltikian, L.~Pryadko and S.~-C.~Zhang, {\em Phys.  Rev.} 
	 {\bf B52}, R8688 (1995); A.~D.~Mirlin, E.~Al'tshuler and 
	 P.~W\"{o}lfle, {\em Annalen der Physik} {\bf 5}, 281 (1996).
\bibitem{experiment} 
	 A.~Geim, S.~Bending and I.~Grigorieva, {\em Phys.  Rev.  Lett.} 
	 {\bf 69}, 2252 (1992); A.~Smith {\em et al.}, {\em Surface 
	 Science} {\bf 361-362}, 349 (1996); A.~Nogaret {\em et al.}, 
	 University of Nottingham preprint (1997).
\bibitem{fluid-dyn} T.~E.~Taber, {\em Fluid Dynamics for Physicists}
	(Cambridge University Press, 1995).  
\bibitem{mancoff} 
	F.~B.~Mancoff {\em et al.} {\em Phys.  Rev.} {\bf B51}, 13269 
	(1995).  
\bibitem{mancoff2} F.~B.~Mancoff {\em et al.} {\em Phys.  
	Rev.} {\bf B53}, R7599 (1996).  
\bibitem{Northrop} T.~G.~Northrop, 
	{\em The Adiabatic Motion of Charged Particles}, (New York, 
	Interscience Publishers, 1963).  
\bibitem{FogShk} M.~M.~Fogler, 
	A.~Yu.~Dobin, V.~I.~Perel, and B.~I.~Shklov\-skii, {\em Phys.  
	Rev.} {\bf B56}, 6823 (1997).  
\bibitem{HedSmith} P.~Hedeg{\aa}rd 
	and A.~Smith, {\em Phys.  Rev.} {\bf B51}, 10869 (1995).  
\bibitem{karen-thesis} K.~Chaltikian, Ph.~D.~Thesis, Stanford 
	University (1997).
\bibitem{PapVar}
	G.~C.~Papanicolaou and S.~Varadhan, in: {\em Boundary Value 
	problems with rapidly oscillating random coefficients}, (North 
	Holland, Amsterdam, 1982), 835.  
\bibitem{McL} D.~McLaughlin, 
	G.~C.~Papanicolaou and O.~Pironneau, {\em SIAM J. Appl.  Math.} 
	{\bf 45}, 780 (1985).  
\bibitem{DrDyk} Yu.~A.~Dreizin and 
	A.~M.~Dykhne, {\em Sov.  Phys.  JETP} {\bf 36}, 127 (1973).  
	\bibitem{Childress} S.~Childress, {\em Phys.  Earth Planet Int.} 
	{\bf 20}, 172 (1979).  
\bibitem{isichenko2} M.~B.~Isichenko {\em 
	et al.}, {\em Sov.  Phys.  JETP} {\bf 69}, 517 (1989).  
	\bibitem{AvMd} M.~Avellaneda and A.~J.~Majda, {\em Comm.  Math.  
	Phys.} {\bf 138}, 339 (1991).  
\bibitem{papanic-rand} A.~Fanjiang 
	and G.~C.~Papanicolaou, Stanford University preprint (1996).
\bibitem{isichenko1} M.~B.~Isichenko, {\em Rev.  Mod.  Phys.}
	{\bf 64}, 961 (1992).  
\bibitem{SalDup} H.~Saleur and 
	B.~Duplantier, {\em Phys.  Rev.  Lett.} {\bf 58}, 2325 (1987).  
\bibitem{gamma}If the probability to have a saddle point at a 
	level in the interval $[B, B+dB ]$ is given by $dP(B) = 
	B^{\gamma}dB$ at small $B$, then $\nu = {4\over 3}(\gamma+1)$.
\end{references}
\end{document}